\documentclass[twocolumn,appendixfloats,tighten]{aastex6}
\usepackage{newtxtext,newtxmath}
\usepackage[T1]{fontenc}
\usepackage{ae,aecompl}

\usepackage{amsmath}	% Advanced maths commands
\usepackage{amssymb}	% Extra maths symbols

\usepackage{ctable}
\usepackage{url}
\usepackage{xspace}
\usepackage[normalem]{ulem}
\usepackage{hyperref}
\usepackage[all]{hypcap}
\usepackage{graphicx}
\def\msun{{\rm\,M_\odot}}

\def\msun{{\rm\,M_\odot}}

\newcommand{\kms}{\, {\rm km\, s}^{-1}}

\newcommand{\be}{\begin{equation}}
\newcommand{\ee}{\end{equation}}

\newcommand{\rsun}{ R_{\odot}}

\def\h2{${\rm\,H_2}$}

\usepackage{color}

\begin{document}

\title{The nearest discovered black hole is likely not in a triple configuration}

\author{Mohammadtaher Safarzadeh\altaffilmark{1}, Silvia Toonen\altaffilmark{2}, Abraham Loeb\altaffilmark{1}}
\altaffiltext{1}{Center for Astrophysics | Harvard \& Smithsonian, 60 Garden Street, Cambridge, MA, USA
\href{mailto:msafarzadeh@cfa.harvard.edu}{msafarzadeh@cfa.harvard.edu}}
\altaffiltext{2}{Institute of Gravitational Wave Astronomy, School of Physics and Astronomy, University of Birmingham, Birmingham, B15 2TT, United Kingdom}

\begin{abstract}
HR 6819 was recently claimed to be a hierarchical triple system of a Be star in a wide orbit around an inner binary system of a black hole (BH) and a B III type star.  We argue that this system is unlikely to be a hierarchical triple due to three reasons: (i) Given that this system is discovered in a magnitude limited Bright Star Catalog, the expected number of such systems in the Milky Way amounts to about $10^4$ while the estimate for the MW budget for such systems is between $10^2-10^3$ systems under generous assumptions.  Such a large gap cannot be reconciled as it would otherwise likely overflow the MW budget for BHs; (ii) The dynamical stability of this system sets lower bounds on the orbital separation of the outer Be star, while it not being resolved by Gaia places an upper limit on its projected sky separation.  We show that these two constraints would imply a narrow range for the outer orbit without resorting to geometrical fine-tuning; (iii) The triple system should have survived the stellar evolution prior to the formation of the BH in the inner binary.  We perform numerical simulations starting with conservative initial conditions of this system and show that a small parameter space for BH progenitor star's mass loss, BH natal kicks, and initial orbital separation can reproduce HR 6819.  Therefore, we propose this system is a chance superposition of a Be star with a binary.  

\end{abstract}

\section{Introduction}\label{sec:intro}
The recent unexpected discoveries of unusual black holes (BHs), either in mass \citep{Liu:2019ch} or in companions \citep{Rivinius:2020gk} provides exciting opportunity to re-visit our assumption regarding the formation of the BHs.
LB-1 was originally claimed to be a 70 $\msun$ BH in a wide orbit around an 8 $\msun$ star \citep{Liu:2019ch}, and more recently HR 6819 is claimed to be a hierarchical triple system with a BH in its inner binary \citep{Rivinius:2020gk}.

While theorists have had a difficult time explaining the formation of LB-1 as a 70 $\msun$ BH in a wide orbit around an 8 $\msun$ star  \citep{2019arXiv191210456S,AbdulMasih:2019vp,ElBadry:2019vd,Eldridge:2019wb}, 
more detailed modelings and observations suggested the system to be a binary of a Be type star and a stripped star \citep{2020arXiv200412882S}.
\citet{Rivinius:2020gk} claims both systems to be hierarchical triples with a Be type star in a wide orbit around an inner binary of a class B star around a BH.      
The lower limits on the mass of the BH in the inner binary is about $\gtrsim 4.2\,(6.3)\,\msun$, and the mass of the B star in the inner binary is found to be $\gtrsim 5.0\,(8.2)\,\msun$ for HR 6819 (LB-1). 

While it is possible to explain the presence of a particular system with certain mass and structure through unconventional channels, three issues are often neglected:
(i) Budget: detecting a system by studying a sample of targets with size $N_t$, implies the presence of  $N_{MW}/N_t$ such systems in the MW where $N_{MW}$ is the expected number of similar targets in the MW. 
One has to check whether the implied number density of such systems is within the allowed range for the MW and if it is higher than the expectations, one has to explain why that is the case;
(ii) Stability: dynamical stability of a system can be perturbed either by the cumulative effect of long-distance encounters of passing by objects or a catastrophic collision with an equal mass object at short distances or internally 
due to the constituents of the system itself, where the last possibility is relevant to the hierarchical triple systems \citep{Mardling:2001dl}; and 
(iii) Lifetime: a short-lived system has a lower detection probability compared to long lived systems.

The structure of this \emph{Letter} is as follows: In \S2 we discuss the inferred budget of the BHs in binaries and triples given the HR 6819 claimed discovery. 
In \S3 we consider limits on the orbital separation of the outer Be star in this system and argue there is only a narrow possible range for this system to a be a hierarchical triple. 
In \S4 we perform numerical simulation of this system from conservative initial conditions showing the survival of such a system needs fine-tuning in initial conditions, 
and in \S5 we summarize our result and discuss the implications. 

\section{MW budget for black holes in triple systems}\label{sec:budget}

HR 6819 is initially discovered in the magnitude limited Bright Star Catalog \citep[BSC; ][]{Hoffleit:1991vr}, which contains about 10,000 stars. 
Of these, there are about 900 early [B0-B2], 600 mid [B3-6], 900 late [B7-9] B-type stars.

The discovery of HR 6819 has been the outcome of studying stars with spectral class similar to the outer Be star in this system. 
Be stars are a subclass of B type stars that are rotating and therefore showing a broad emission line in their spectra which constitute about 10\% of all B-type stars.

To estimate the expected number of B-type stars in the MW, we start with the formation rate of 8-20 $\msun$ stars, for which we adopt a value $R_8\approx0.02$ yr$^{-1}$ \citep{Licquia:2015eb}. 
The formation rate of other classes of stars can be rescaled given their mass. 
Assuming a Salpeter initial mass function ($dN/dM\propto M^{-2.35}$), B-type star with mass $>3~\msun$ form at a rate $R_3\approx4\times R_8$. 
For a B-III type star, similar to the star in the inner orbit of HR 6819 with mass $M\approx5~\msun$, we arrive at $R_{B3}\approx2\times R_8$. 
The expected number of a class of B3 type stars in the MW can be computed as:
\be
N_{B3}\approx R_{B3} \times t_{\rm ms},
\ee
where $t_{\rm ms}$ is the main sequence (MS) lifetime of a star \citep[$t_{\rm ms}\propto m^{-2.92}$;][]{Demircan:1991ee}.
Given that the lifetime of $8 \msun$ stars is approximately 50 Myr \citep{Cummings:2018cz}, the main sequence age of a 5 $\msun$ star is about 200 Myr, and therefore, the expected number of B3-type stars in the MW amount to $N_{B3}\approx8\times10^6$.

%Given that in the BSC sample, there are about 600 B3 type stars, one can, therefore, conclude that the expected number of HR 6819 type systems in the MW is 
%$N_{\rm HR 6819}^{\rm obs}\approx1.3\times10^4$.
If we assume that the search strategy for this systems was motivated by looking at the Be type stars in the sample, given that about 100 such stars have been studied, and assuming Be stars constitute about 10\% of all B-type stars, 
the expected number of HR 6819 type systems in the MW amounts to $N_{\rm HR 6819}^{\rm obs}\approx8\times10^3$.

The numbers from the detection frequency of the HR 6819 should be checked against the expected MW budget from other lines of evidence:
The total number of BHs in triple configurations can be estimated as:
\be
N_{\rm HR 6819}^{\rm theory}= f_{\rm B3}\times f_{\rm Be} \times N_{\rm BH,triple}, 
\ee
where 
\be
N_{\rm BH,triple}=f_{\rm 3/2} \times N_{\rm BH,ms}.
\ee
Here $f_{\rm 3/2}$ indicates the relative ratio of all systems in triple configuration to those in binary which ranges from $\approx1.5-2$ for stars with masses in the range of $20-40~\msun$ \citep{Moe:2017eg}.
$f_{\rm B3}$, and $f_{\rm Be}$ indicate the fraction of triples with a B3 star in the inner orbit, and a Be star in the outer orbit, and $N_{\rm BH,ms}$ indicates the number of BHs in binaries around main sequence (MS) stars.
Conservatively, we assume a flat probability distribution for the mass of the star in the inner binary and the outer binary, with maximum mass progenitors being less than the mass of the BH progenitor star.
The probability of having a certain spectral class in the triple scales with the lifetime of that type. 
Therefore, $f_{\rm B3}\approx t_{5}/t_{\rm MW}\approx 10^{-2}$, where $t_{5}$ is the main sequence lifetime of a 5 $\msun$ star, and we consider $t_{\rm MW}=10$ Gyr.
As only 10\% of B type stars are found to be in the Be spectral class, we adopt $f_{\rm Be}\approx 0.1$. Therefore the expected $N_{\rm HR 6819}^{\rm theory}\approx 2\times10^{-3} N_{\rm BH,ms}$.
If we assume that the Be star's presence is related with being in such a hierarchical systems, (i.e., $f_{\rm Be}\approx1$), we arrive at the expected $N_{\rm HR 6819}^{\rm theory}\approx 2\times10^{-2} N_{\rm BH,ms}$.

Through a series of population synthesis of binaries in the MW, \citet{Olejak:2019wj} concludes there are $\approx 10^7$ BHs in binary systems in the MW with average BH mass of about 20 $\msun$. 
This includes all forms of binaries, including BH-BH, BH-Neutron star, BH-White Dwarf, and BH-MS. 
For example, a separate study by \citet{Lamberts:2018jw} arrives at $\approx1.2\times10^6$ binary black holes in the MW with the mean mass of 28 $\msun$, which is a subset of the all BHs in binaries.
The relevant budget to our case is the total estimated number of BH-MS systems which is estimated to be $3.4\times10^4$, $1.2\times10^5$, and $1.2\times10^4$ in the bulge, thin disk, and thick disk respectively. 
This amounts to $N_{\rm BH,ms}\approx 10^5$ BH-MS binaries in the MW. 
Combining the above we arrive at an expected number of $N_{\rm HR 6819}^{\rm theory} \approx 10^2-10^3$ in the MW, depending on whether we consider the likelihood of having a Be star as independent of it being in triple configuration or not.
We note that we have ignored the impact of flat IMF assumption for the two stars in this calculation to be on the extremely conservative side. 

The discrepancy between the implied number of HR 6819 type systems in the MW from the observations and the expected number from theory is about 1 to 2 orders of magnitude part of which we might be able to account for by changing parameters in the population synthesis codes. However, the total expected BHs in the MW is estimated to be around $10^8$ \citep{Shapiro:2008ke}, 
and about $10^9$ based on microlensing studies \citep{Agol:2002ii}. The required boost in the expected number of BHs in binaries would mean a considerable fraction of the BHs of the MW is locked up in binaries, which is unlikely. 

At this point we should mention an important caveat: The budget estimate presented in this section is the weakest of the three arguments we lay out against HR 6819 being a triple system.
One major obstacle is that any population of stars of a specific spectral class/surface temperature will be a combination of varying stellar ages and initial masses. 
Therefore, one needs to model this by a population synthesis code to arrive at a more accurate estimate, which is beyond the scope of our \emph{Letter}.
However, even the populations synthesis codes are extremely uncertain. 
For example, the low BH mass in HR 6819 is not a typical BH mass in \citet{Olejak:2019wj}, and only specific codes can generate such systems \citep{Eldridge:2019wb}. 
Moreover, formation of HR 6819 might have involved many possible interacting binary stellar evolutionary pathways, which makes a budget estimate extremely difficult. 
Therefore, we encourage the readers to take these numbers with caution.

\section{Constraints on the outer orbital separation}\label{sec:triple}
The bounds on the orbital separation of the Be star can be divided into lower and upper bounds. 
The lower bounds come from the fact that the system is stable: if the system becomes unstable the orbits of the stars change on the dynamical timescale, and most likely the interaction leads to a dissolution 
\citep[e.g., ][]{vandenBerk:2007ju} and it is unlikely we're observing the system during such interaction since the inner orbit shows stable periodicity and low eccentricity. 
The upper bounds come from the fact that the system is not resolved in the Gaia mission, which we will discuss towards the end of this section. 

\subsection{Lower bound}

Survival of hierarchical triple systems relies on certain ratios to be in place with regards to the relative ratio of the 
semi-major axis of the outer orbit ($a_2$) to the inner binary's ($a_1$). 
These ratios set both the dynamical stability and the induced oscillations on to the inner binary due to the presence of a third body. 

\subsubsection{Stability Criteria}

\begin{figure}
 \includegraphics[width=0.9\columnwidth]{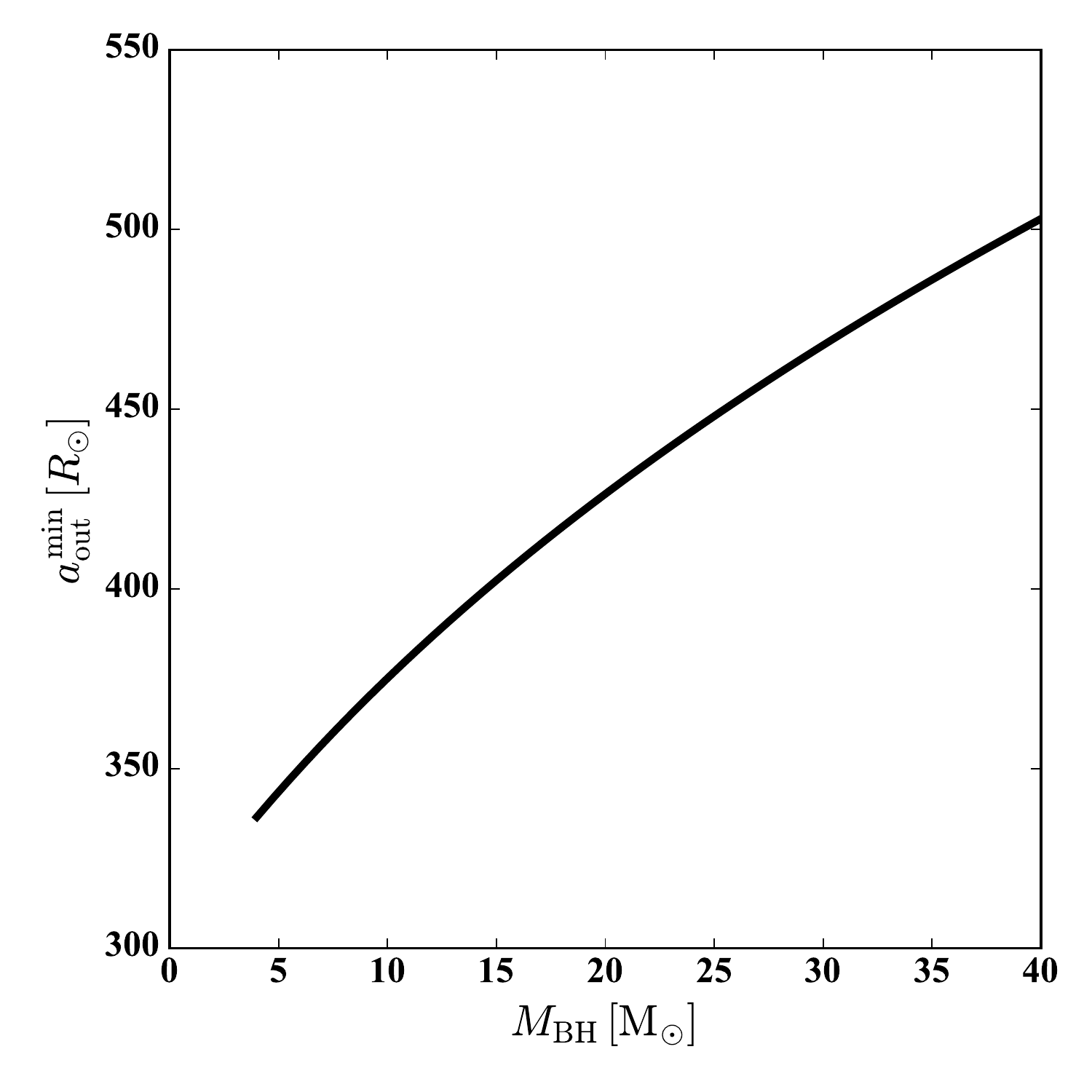}
 \caption{The minimum separation of the outer Be star from the inner binary consisting of a BH of mass $M_{BH}$ and a B III type star with mass $5~\msun$ assuming the inner orbital separation is 100 $\rsun$.
 The bound is due to requiring the triple system to remain dynamically stable following \citet{Mardling:2001dl}.}
 \label{fig:R_p_min}
\end{figure}

We implement the stability criteria for a triple system following \citet{Mardling:2001dl}:
\be \label{eq:stab}
a_{\rm out}^{\rm min}=C \left[ (1+q_{\rm out})\frac{1+e_{\rm out}}{(1-e_{\rm out})^{1/2}} \right]^{0.4} a_{\rm in},
\ee
where $m_1$, and $m_2$ are the masses of the objects in the inner binary, and $m_3$ is the mass of the third outer body. 
Here, $q_{\rm out}=m_3/(m_1+m_2)$, and $e_{\rm out}$ is the eccentricity of the outer orbit.  The value of the constant is empirically found to be $C=2.8$. 
For a given $e_{\rm out}$, and $q_{\rm out}$, the stability of a triple configuration would require that
$a_{\rm out}>a_{\rm out}^{\rm min}$.
Since only a lower limit of $M_{BH}=4~\msun$ is reported for HR 6819, the outer radius would be a function of the inner BH mass. 
Figure \ref{fig:R_p_min} shows $a_{\rm out}^{\rm min}$ as a function of the inner BH mass. If the BH mass is 4 $\msun$,  $a_{\rm out}^{\rm min}\approx350~\rsun$, while
if the inner BH is $40~\msun$, $a_{\rm out}$ should be larger than $\approx500~\rsun$.

We note that we have assumed $e_{\rm out}\approx0$, which is a conservative assumption if we want to require the triple system to be in a hierarchical structure for long term stability as otherwise $m_3$ can come close to the inner binary and perturb the system. 
Assuming higher outer eccentricity would mean larger outer orbital separation for dynamical stability.

\subsubsection{Lidov-Kozai timescale}
In the presence of a third body, the inner orbit will experience eccentricity cycles known as Lidov-Kozai \citep[LK; ][]{Lidov:1962du,Kozai:1962bo} cycles. 
We require this timescale to be longer than the age of the system as otherwise the star and the BH in the inner orbit would have interacted, leading to mass transfer and shrinkage of the inner orbit, and perhaps merging. 
The LK timescale is given by \citep[e.g., ][]{Toonen:2016tw}:

\be
\label{eq:kozai_timescale}
t_{\textrm{LK}} \approx \frac{P_{\rm out}^2}{P_{\rm in}}\frac{m_1+m_2+m_3}{m_3} \left(1-e_{\rm out}^2\right)^{3/2}
\ee

\begin{figure}
 \includegraphics[width=0.9\columnwidth]{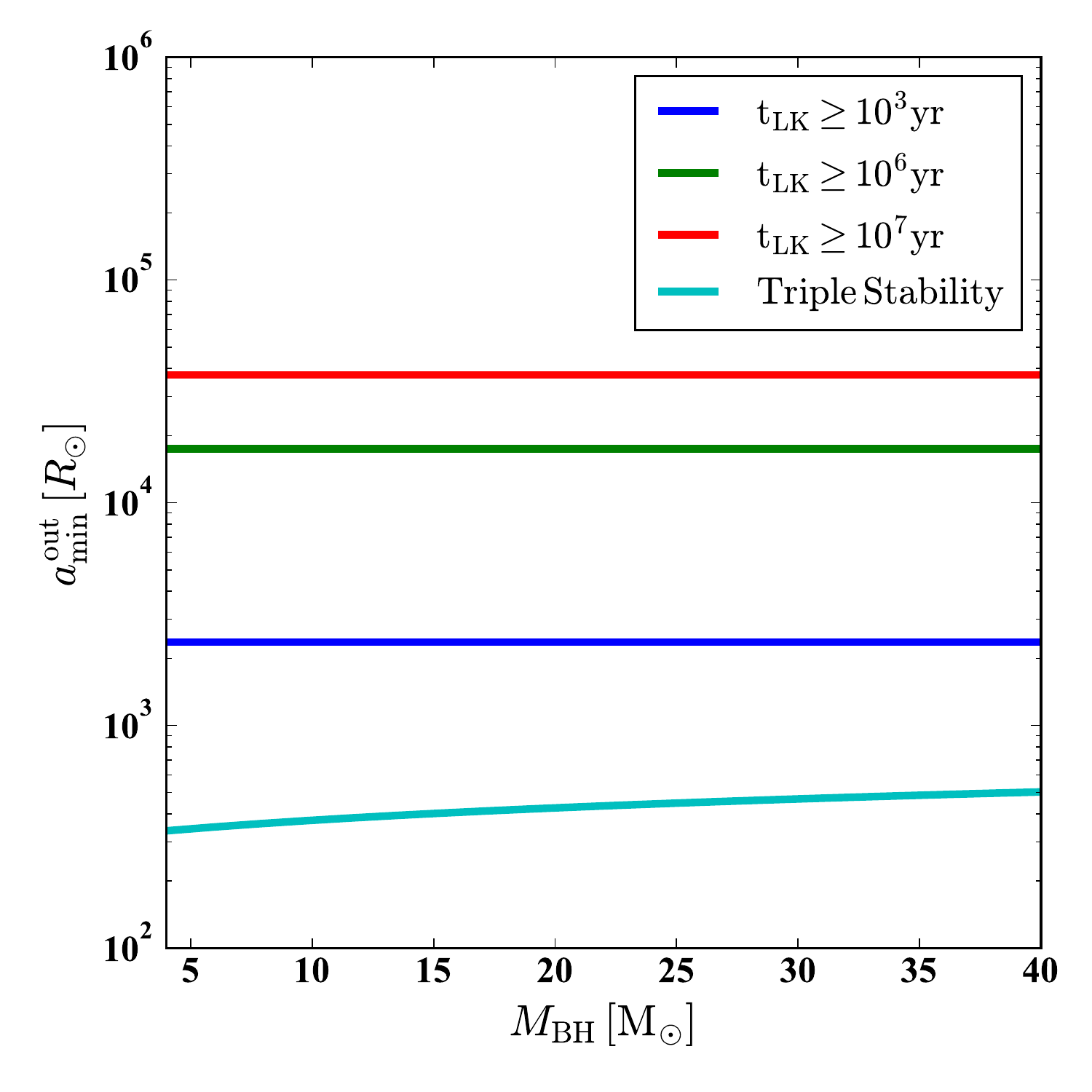}
 \caption{The minimum outer orbit separation requiring the induced LK timescale to be longer than the age of the system which we assume different values for, as a function of the inner BH mass. 
 Over-plotted in blue is the bound from the dynamical stability argument. 
 The bounds from LK timescale is stronger than dynamical stability arguments. We note that this is the absolute minimum timescale one can adopt for this system.}
 \label{fig:comb}
\end{figure}

For each inner BH mass, we can set the lower limit on $a_{\rm min}^{\rm out}$ to give an LK induced timescale comparable to the age of the system. 
If we assume an age of 1 Myr, the outer binary separation should be $a_{\rm out}\gtrsim10^4~\rsun$, weakly dependent on the inner 
BH mass, as shown in Figure \ref{fig:comb}. If we adopt such a line of reasoning, we arrive at $a_{\rm min}^{\rm out}\approx10^4~\rsun$ if the age of the system is about 1 Myr. 

We don't know the age of the system other than it should be less than the $t_{ms}$ of the B III star in the inner orbit ($\approx$200 Myr). 
One way out would be that the system is co-planar in all components and, therefore, there are no LK induced oscillations. 
The co-planar configuration can arise if the system is formed from a disk. However, \citet{Tokovinin:2017jy} found that for triples with massive primaries or large outer orbit separations, the angle between the inner and outer orbit becomes more randomly oriented. 
For randomly distributed angles between the inner and outer orbits there is a 78\% chance to have the angle lie between $39^\circ\leq \Phi \leq 141^\circ$ which would lead to LK cycles. 

We note that the LK cycles could be damped due to astrophysical processes in the inner orbit that dominate over the apsidal precession from the outer body such as tides. 
However, such processes would only take place when the inner binary separation is about a few stellar radii \citep{Fabrycky:2007jh}, which is far smaller than the case for HR 6819.

\subsection{Upper bound}
The very fact that the outer binary is not resolved in the Gaia mission sets an upper limit on $a_{\rm out}$. The critical angular separation in arcsec would depend on the G band 
magnitude difference between two stars in a binary \citep{deBruijne:2015iu,Toonen:2017gp}:
\be
\rm log(s_{\rm crit,gaia}) = 0.075\mid\Delta G\mid -0.53.
\ee
Assuming $\mid\Delta G\mid=0$, and a distance of 300 pc, we get an upper limit on the projected sky separation of $d\lesssim20,000~\rsun$. The projected sky separation is related to the outer orbital separation as $d=a_{\rm out}\times \sin~(\theta)$, 
where $\theta$ is the angle between the two vectors, one from the inner binary to the observer and the other from the inner binary to the outer Be star. 
Given the mild dependence of the upper bound on the G band mag difference, 
one can conclude that the outer Be star orbital separation should be less than $2\times10^{4}~\rsun$ without considering the projection effects. However, to allow for larger separations, we have to confine $\theta$ to smaller angles. For example, it is possible to have a very large $a_{\rm out}$ if the Be star lies along the line of sight to the binary. One can compute the probability of such configuration by $p(\theta)=S(\theta)/4\pi$ where $S(\theta)$ is the solid angle corresponding to $\theta$ in radians. 
For example, the probability of $a_{\rm out}\gtrsim2\times10^5~\rsun$, is less than about 1\%. 

\section{Numerical simulation}
Separate from the constraints on the outer orbit from the present-day configuration of the system, HR 6819 should have remained bound throughout the stellar evolutionary phases of the progenitor star that made the inner orbit BH. 
To study this, we perform numerical simulations of a triple system from conservative initial conditions, such that the triple starts in a bound configuration. 

We note that in this section we assume that the inner orbit is circularized due to either mass transfer or tides between the black hole and the B star.
We assume that the configuration of the triple after the progenitor of the final BH has gone through a mass transfer phase is $a_{\rm in}=25~\rsun$, $a_{\rm out}=250~\rsun$ with zero eccentricity.
We assumed circular and co-planar orbits since the outer orbit is relatively compact, and for those orbits, \citet{Tokovinin:2017jy} found a tendency towards co-planar orbits. 
The masses of the two stars in the inner and outer orbits are assumed to be 5 $\msun$. 
We also note that our choice of co-planar configuration is also driven by results from the previous section that we concluded a co-planar setup is preferred for the system to avoid LK cycles. 

For the progenitor star of the BH before SNe explosion, we assume a range of masses with the final BH mass always at 5 $\msun$. 
Therefore, the final configuration is determined by two factors: (i) how much mass is lost in the SNe explosion, and (ii) whether there was a natal kick imparted to the BH at formation. 
We compute the fraction of survived systems ($p_{\rm surv}$) after 10,000 trials. 
Survived systems are defined to have bound inner and outer orbits and to be dynamically stable. 
Of all those that have survived, only a fraction would look similar to HR 6819 in terms of the inner orbital parameters and the minimum separation of the outer orbits due to limitations on the LK timescale. 
We note that we do not provide an analytical estimate for survival of triple systems, however, such analytic estimates has been formulated before \citep{Tauris:2017cf}, and implemented in \citep{2019arXiv191210456S} as an example.

\subsection{No kick scenario}
If we assume the BH is born with no natal kick, the final configuration is solely determined by the amount of mass lost in the SNe explosion ($M_{\rm ej}$). 
The ejected mass predicts a unique value for the eccentricity of the inner orbit ($e_{\rm in}$), and depending on the phase of the orbits, it will result in a range of separation and eccentricity for the outer orbit. 
The survival fraction is $p_{\rm surv}= 0\%$ for simulations with $M_{\rm ej}=10~\msun$. The survived fraction remains at 0\% had we assumed smaller $a_{\rm in}=5~\rsun$.
For simulations with $M_{\rm ej}=3~\msun$, initial $a_{\rm in}=5\rsun$, and $a_{\rm out}=50\rsun$; we obtain a $p_{\rm surv}= 52\%$, where about 34\% result in dynamically unstable configurations, and about 14\% having their outer binary disrupted.
However, for those that survive, the inner orbital eccentricity of the survived systems is $e_{\rm in}\approx0.3$ which is 
ten times larger than the upper limit reported for HR 6819, $e_{\rm in}^{\rm HR 6819}<0.03$. This result on the inner orbit eccentricity is independent of the assumed initial inner binary separation. 

To obtain low eccentricities for the inner orbit consistent with HR 6819 we need $M_{\rm ej}<0.3~\msun$ and an initial inner binary orbital separation of $a_{\rm in}\approx100~\rsun$. However, this is basically assuming 
the initial configuration of the systems is similar to its current configuration requiring fine-tuning.  

In summary, if the amount of ejected mass is large, the survived fraction is nearly zero.
 For small ejected masses, the inner orbit will have a large eccentricity compared to the value for HR 6819 ($e_{\rm in}^{\rm HR 6819}<0.03$). 

\subsection{With BH kicks}
The impact of kicks on hierarchical triples has been investigated before \citep{Pijloo:2012kw}.
For simulations with large ejected mass ($M_{\rm ej}=10~\msun$), which are already unbound due to mass loss alone, the addition of BH natal kick does not help to increase the survived fraction. 
Simulations with a smaller ejected mass and no natal kicks resulted in large eccentricities for the inner binary. 
Here we explore whether imparting BH natal kick would help to account for the inner orbit characteristics of HR 6819.

For simulations with $M_{\rm ej}=3~\msun$, $a_{\rm in}=25\rsun$, and initial $a_{\rm out}=250\rsun$, the survived fraction drops from 52\% in the case of no kicks, to 25\%, 10\%, and 3\% for the simulation with BH natal kick with magnitudes of 50, 100, and 150 $\kms$.
For the same simulations but with $a_{\rm in}=5\rsun$, and initial $a_{\rm out}=50\rsun$, the survived fraction drops from 50\% in the case of no kicks, to about 45\%, 27\%, and 20\% for the simulation with BH natal kick with magnitudes of 50, 100, and 150 $\kms$. We note that depending on the direction of the imparted kicks, it is possible to either result in the system to remain bound or to disrupt the system.

\begin{figure}
 \includegraphics[width=0.9\columnwidth]{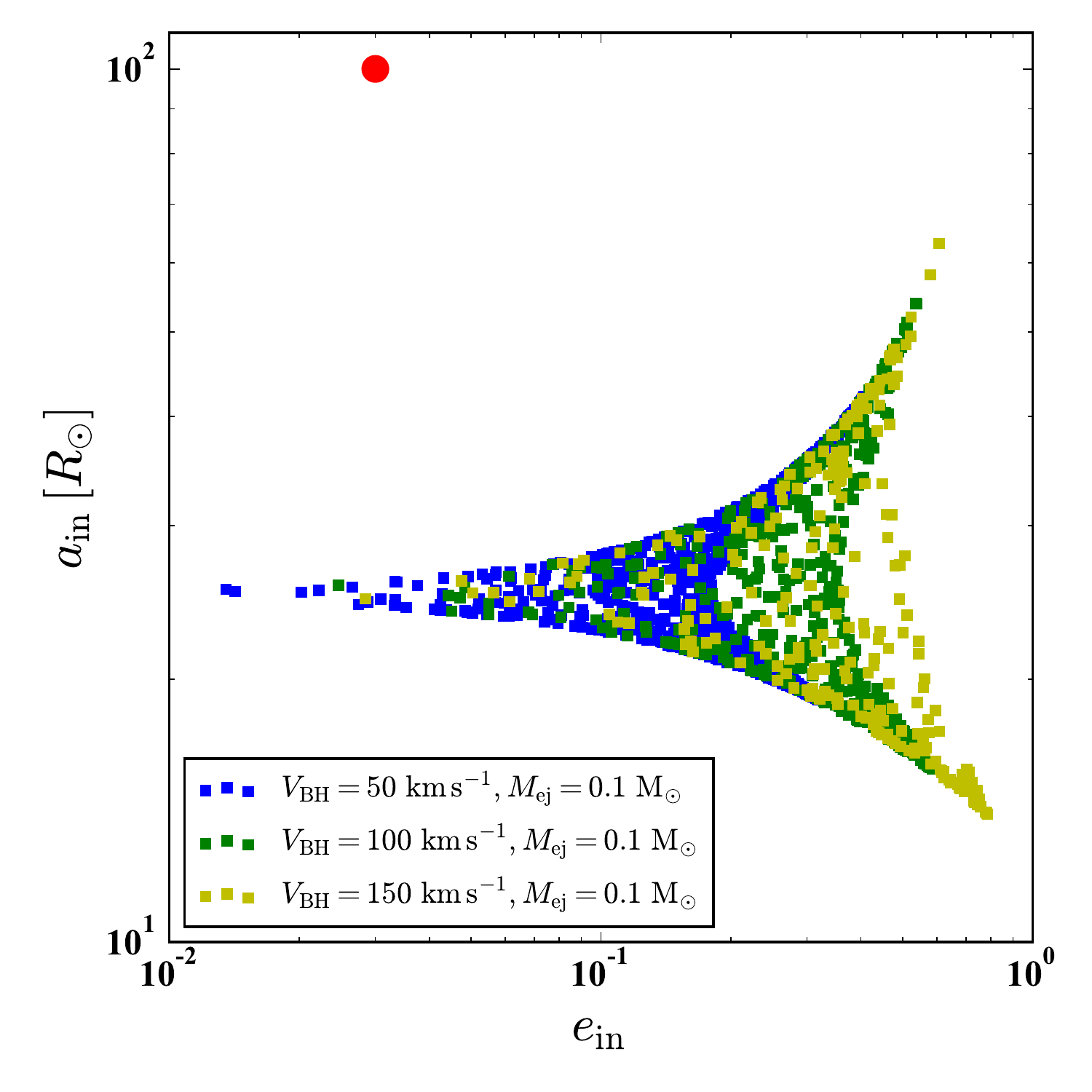}
  \includegraphics[width=0.9\columnwidth]{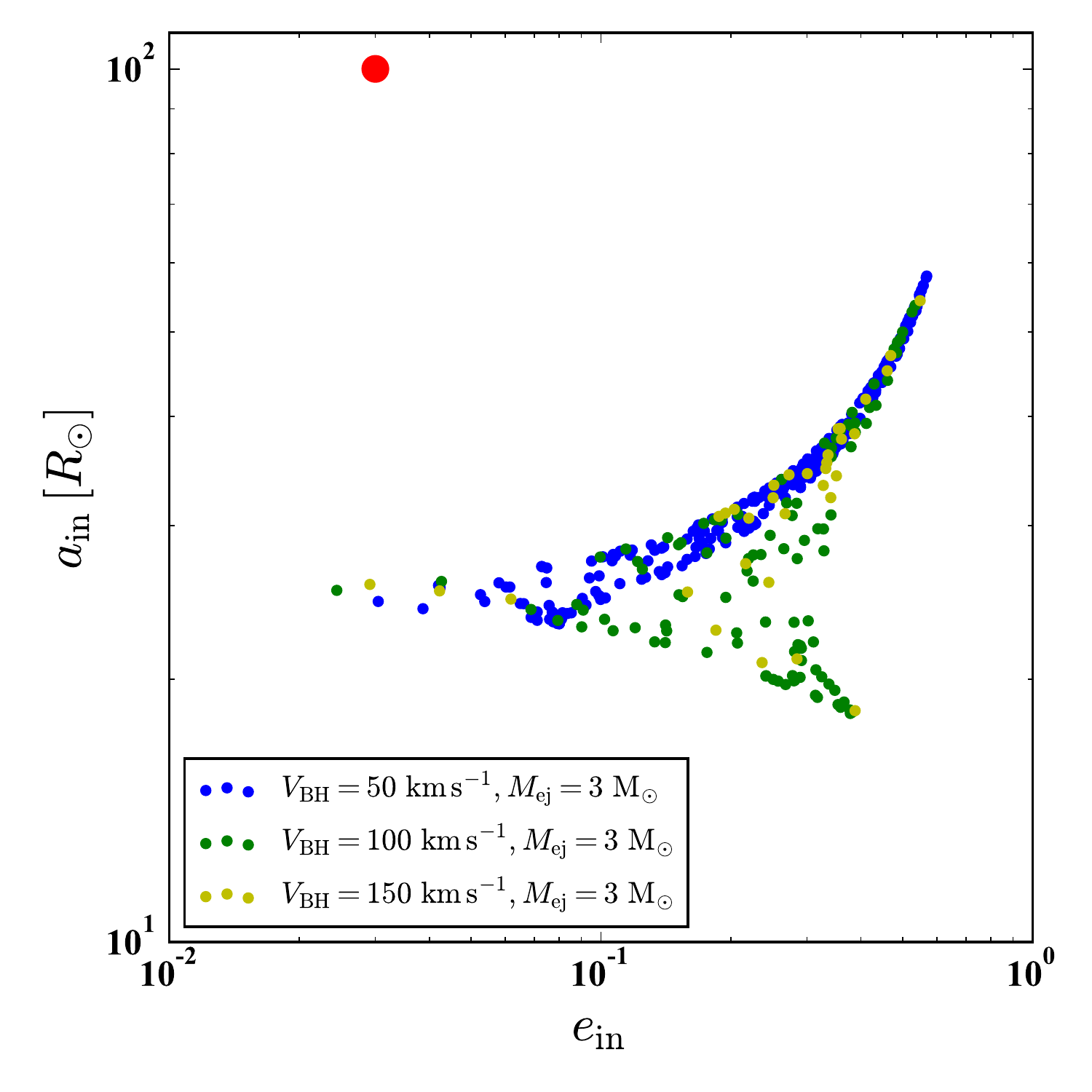}
 \caption{\emph{Top panel:} results from a simulation with $M_{\rm ej}=0.1~\msun$ for various BH natal kicks. \emph{Bottom panel:} the same for the simulations with  $M_{\rm ej}=3~\msun$. 
The red dot in both plots indicates the location of HR6918 in the eccentricity-orbital separation of the inner orbit. The surviving systems never come close to HR 6819 except with fine-tuning the initial conditions.}
 \label{fig:sims}
\end{figure}

Figure \ref{fig:sims} shows the final eccentricity and orbital separation of the inner binary from such simulations assuming the outer orbital separation is initially set at twice the minimum separation required for dynamical stability. 
In the top panel, we show the result of a simulation with $M_{\rm ej}=0.1~\msun$ for various BH natal kicks, and in the bottom panel, we show the same for the simulations with  $M_{\rm ej}=3~\msun$. 
The red dot in both plots indicates the location of HR6918 in the eccentricity-orbital separation of the inner orbit. For our adopted initial configuration our survived systems never come close to HR 6819.

If we perform similar simulations with larger inner orbit separation, such as $a_{\rm in}=90~\rsun$, still only a tiny fraction (close to 1\%) of the simulations would resemble the inner orbit characteristics of HR 6819.
Therefore, one can conclude that similar to the previous section, a fine-tuning of initial conditions would be required to match HR 6819. 

We note that there are other stellar evolutionary physics that we have not modeled in our work, such as the possibility of mass transfer during an LK induced oscillation. 
While it is unclear whether such  processes help the survival of the system or not, we leave this as a caveat of our work, which we will return to with detailed modeling in a future work.

\section{summary and discussion}\label{sec:sum}
HR 6819 is the closest known BH  and argued to be in a triple configuration with an outer Be star. 
We have shown that such a configuration is likely not viable for this system due to two reasons: (i) Given an average lifetime of B3 and Be stars (less than about 200 Myr), such systems must be born within the
last few hundred million years. The detection frequency of one HR 6819 system in 600 B3 type stars in the BSC implies about $10^4$ such systems should be residing in the MW. 
However, conservative theoretical expectations would predict at most between $10^2-10^3$ such systems. This large discrepancy can is not easy to be reconciled without overflowing the MW budget for BHs, and  
(ii) This system is not resolved in the Gaia dataset, which sets an upper limit on the projected sky separation of its outer orbit ($a_{\rm out}$). Together with requiring dynamical stability for the system 
these considerations tighten the allowed range for $a_{\rm out}$ to within almost 1 dex implying that fine-tuning of $a_{\rm out}$ would be needed for this system to survive,
If we assume the system is not co-planar in all components, one requires the Lidov-Kozai induced oscillation timescale to be longer than the observed timescale of the system as otherwise such oscillations should have been detected in the data.
Such considerations would tighten the allowed range further. For example, assuming the LK timescale to be on the order of 100 Myr (age of the system) would imply an extreme fine turning for the geometry of the system. 
(iii) This system should have survived the stellar evolution that has taken place before the formation of the inner BH. 
Given that the inner binary has a BH of a mass at least 5 $\msun$, we simulated the final configuration of this system for a set of conservative initial conditions. 
We varied the ejected mass during the formation of the inner BH from its pre-SNe progenitor star, and examined different natal kick magnitudes imparted to the BH at birth with random directions. 
We have shown that either the triple configuration is not stable and the system dissolves, 
or the inner orbit configuration does not resemble the rather circular inner orbit of the HR 6819. 

Can we conclude that HR 6819 is a binary of a BH with a B3 type star, and the outer Be star is chance superposition along the line of sight to the system?
The expected number of Be type stars in the MW is about $\approx3.6\times10^7$. Adopting the fiducial values of for the MW disk radii (8.5 kpc) and scale height for young stars (100 pc), 
one can compute the average density of such stars in the MW. 
On the other hand, if we assume the presence of the Be star is a chance alignment, 
the projected sky separation between the B III and Be star should be less than the Gaia resolution (0.3 arcsec for stars with similar G band magnitude). 
One can compute the volume within the geometrical cone of the Gaia resolution and a depth of 2 kpc leads and assuming 0.3 arcsec projected sky separation for each B III star.
Given about 3000 B III star targets in the BSC, this leads to a 10\% chance alignment under conservative estimates. 
This order of magnitude calculation shows that the chance alignment of a B III and Be star within the Gaia resolution might be a potential explanation for this system. 

In this work, we have attempted to provide an order of magnitude approach to the question at hand without getting into detailed numerical work. 
However, all of the three arguments that we provided support each other in the fact that a triple interpretation of HR 6819 is in tension with theoretical expectations. 
While we have focused our work on one particular system, the approach presented here applies to all triples and higher multiple structures that are detected in the MW.

\acknowledgements We are thankful to the referee for their detailed comments which improved the presentation of our work. We thank Thomas Rivinius, and Adrian Hamers for useful discussions. 
This work is supported by the National Science Foundation under Grant No. AST-1440254, and by Harvard's Black Hole Initiative, which is funded by JTF and GBMF. 
ST acknowledges support from the Netherlands Research Council NWO (VENI 639.041.645 grants).

\bibliographystyle{yahapj}
\bibliography{the_entire_lib.bib}
\end{document}